\documentclass[11pt,twoside]{article}
\usepackage{asp2010}

\resetcounters

\begin{document}

\title{Determining the forsterite abundance of the dust around AGB stars}
\author{B.L. de Vries$^{1}$, M. Min$^{2, 3}$, L.B.F.M. Waters$^{4,3}$, J.A.D.L. Blommaert$^1$, F. Kemper$^5$
\affil{$^1$Instituut voor Sterrenkunde, K.U. Leuven, Celestijnenlaan 200D, 3001 Leuven, Belgium, e-mail: ben@ster.kuleuven.be }
\affil{$^2$Astronomical Institute, University of Utrecht, Utrecht, The Netherlands    }
\affil{$^3$Sterrenkundig Instituut Anton Pannekoek, University of Amsterdam, Science Park 904, 1098 XH, Amsterdam, The Netherlands    }
\affil{$^4$Space Research Organization of the Netherlands (SRON) 9700 AV Groningen, Netherlands}
\affil{$^5$Jodrell Bank Centre for Astrophysics, Alan Turing Building, School of Physics and Astronomy, University of Manchester,
Oxford Road, Manchester, M13 9PL, UK}
}

\begin{abstract}
We present a diagnostic tool to determine the forsterite abundance of the dust ejected by AGB stars. Our method is based on a comparison between the observed strength of spectral bands of forsterite and model calculations. We show that the 11.3 $\mu$m forsterite band is a robust indicator of the forsterite abundance of the current mass-loss period for AGB stars with an optically thick dust shell. The 33.6$\, \mu$m band of forsterite is sensitive to changes in the density and the geometry of the emitting dust shell, and so a less robust indicator. We apply this method to six high mass-loss rate AGB stars, showing that AGB stars can have forsterite abundances of 12\% by mass and higher, which is more than the previously found maximum abundance of 5\%.
\end{abstract}

\section{Introduction}
One of the most important aspects of AGB stars is their stellar wind. AGB stars can have mass-loss rates between $10^{-8}$ \nolinebreak M$_{\odot}$ \nolinebreak yr$^{-1}$ and about $10^{-4}$ M$_{\odot}$ yr$^{-1}$. The high mass-loss rate automatically constrains the life-time of these objects, and gives rise to dense circumstellar envelopes. When the gas in this outflow reaches distances where the temperature is low enough the material can condense into solid particles (from here on called dust). 

Even though amorphous silicate dust is more abundant, crystalline silicate dust is also seen in many astronomical environments like disks around pre-main-sequence stars \citep{waelkens96, meeus01,spitzer1}, comets \citep{wooden02}, post-main-sequence stars \citep{waters96, syl99, mol02} and active galaxies \citep{kemper07, spoon06}. Theoretical studies predict that the formation of crystalline silicate dust is favored at high gas densities \citep{tielens98,gailsedl99, sogawa99}. \citet{speck08} also argue that there is a correlation between crystallinity, mass-loss rate and the initial mass of the star.

These suggestions are consistent with the detection of crystalline silicates in the high density winds of OH/IR stars. Indeed, observations show a lack of crystalline silicate dust bands in the spectra of low mass-loss rate AGB stars \citep{waters96, syl99}.
However, an alternative explanation for this lack exists. \citet{kemper01} have shown that a contrast effect hinders the detection of crystalline silicate bands in infrared spectra of low mass-loss rate AGB stars.

\citet{kemper01} obtained an indication for the crystalline silicate abundance in the outflow of AGB stars by comparing the strength of crystalline silicate emission bands of observations to those of model spectra. For the crystalline silicates enstatite and forsterite they measured a maximum abundance of $\sim$5\% by mass. 

AGB stars are the main contributors of crystalline silicate material to the ISM. The creation of crystalline silicates in the ISM is not possible since the temperatures are too low. If we assume that crystalline silicates are not destroyed in the ISM, we expect their abundance in the diffuse ISM to be comparable to that in AGB winds. The crystalline silicate abundance in the diffuse ISM has been investigated by \citet{kemper04, kemperErr05}. They looked at the line of sight towards the Galactic Centre and found an upper limit of the degree of crystallinity in the diffuse ISM of 2.2\% by mass. The abundance found in AGB stars by \citet{kemper01} is therefore higher than that for the diffuse ISM, suggesting the crystalline material may be effectively destroyed or hidden. Possible mechanisms could be amorphization by inclusion of iron atoms in the crystalline lattice, shocks and UV radiation (see \citet{kemper04, kemperErr05} for a discussion of such mechanisms).

Determining the crystalline fraction of silicates in AGB stars, as a function of mass-loss rate, will enhance our understanding in two areas: First, what are the conditions needed for the formation of crystalline silicates? Dust condensation theory suggests that the density is critical in the formation of crystalline silicates \citep{tielens98,gailsedl99, sogawa99}, but this has still not been verified by observations. Second, how much crystalline silicate material is deposited by AGB stars to the diffuse ISM? Do the abundances in the outflows of AGB stars and the ISM match?

In this article we discuss a method to measure the forsterite abundance in AGB stars as applied by \citet{kemper01} and expanded upon by \citet{devries10}.

\begin{figure}
\begin{center}
\includegraphics[scale = 0.6]{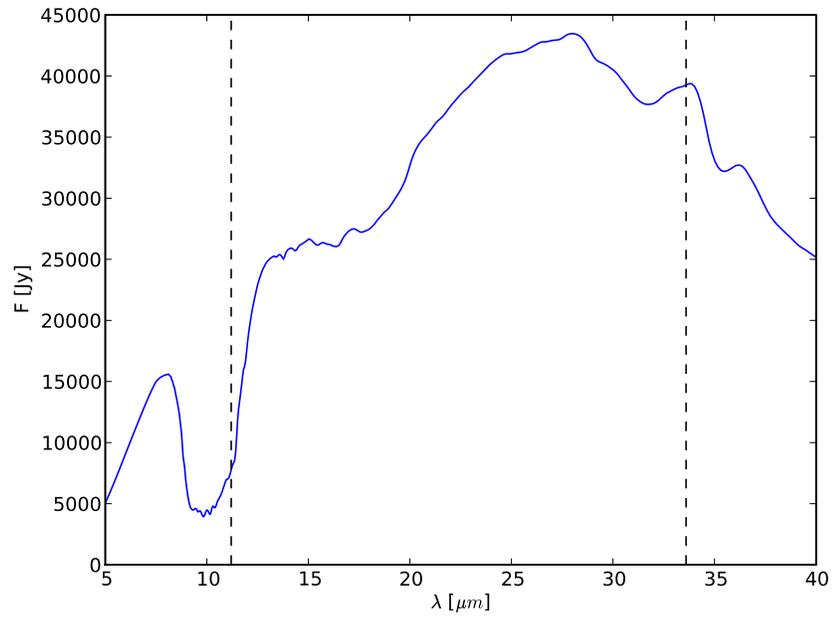} 
\end{center}
\caption{A synthetic spectrum of an AGB star with a mass-loss rate of $3\times10^{-5}$ M$_{\odot}$ yr$^{-1}$. The two dashed vertical lines indicate the position of the 11.3$\, \mu$m and 33.6$\, \mu$m band of forsterite.}
\label{totalSpec}
\end{figure}

\begin{figure}
\begin{center}
\includegraphics[scale = 0.6]{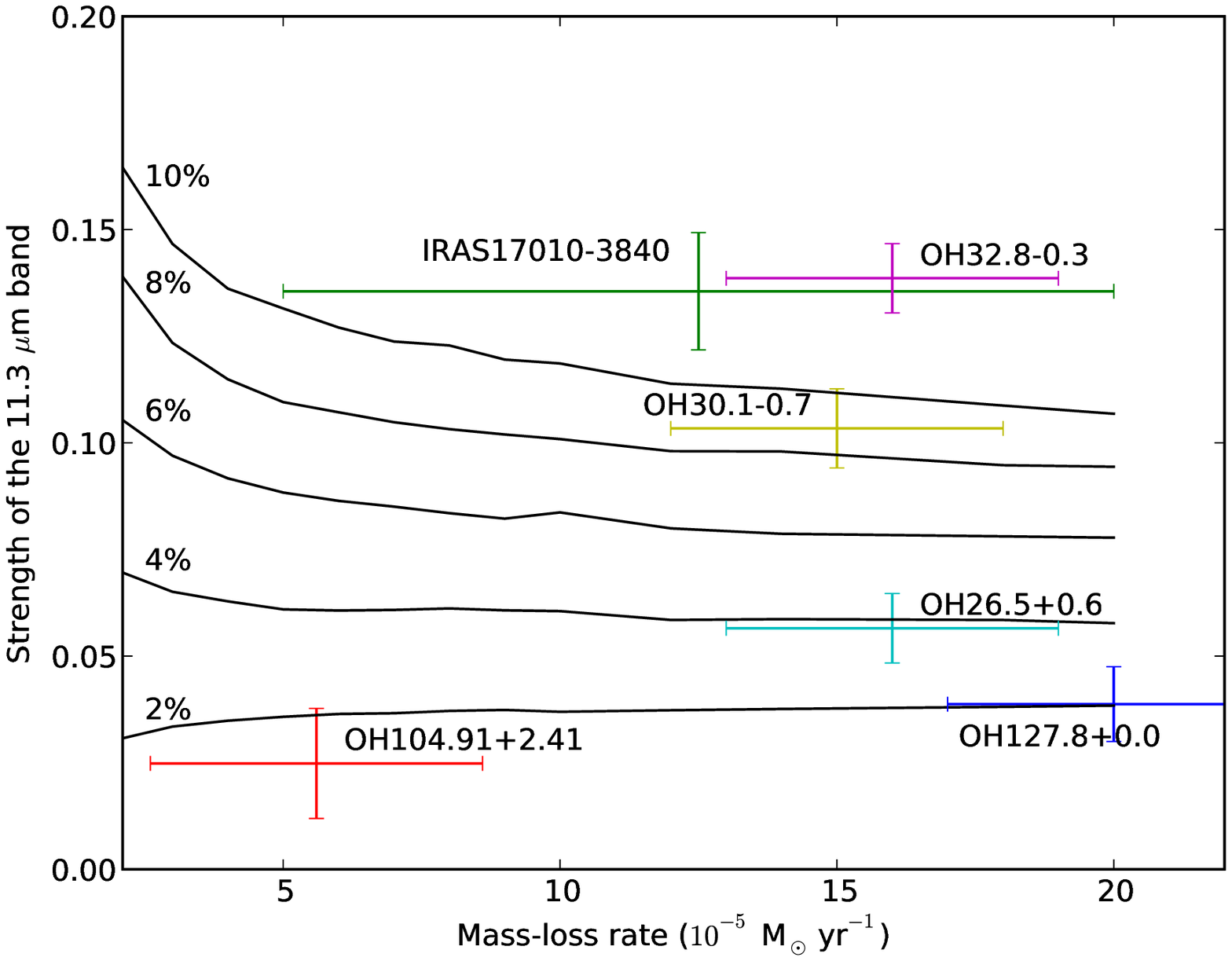}
\end{center}
\caption{The equivalent width of the 11.3$\, \mu$m band for models as a function of the mass-loss rate (lines). The different curves have different forsterite abundances. Also the strength of the feature is shown for six AGB stars observed with ISO-SWS (points with error bars). Figure taken from \citet{devries10}. }
\label{res}
\end{figure}

\section{Using spectral features to measure the forsterite abundance}
One strategy to obtain an indication of the abundance of a dust species is to compare the strength of spectral features from model spectra with those of observed spectra. The usefulness of this method is that once the models have been produced, everyone can easily use them for comparison with their observations. For the crystalline olivine forsterite this method has been used by \citet{kemper01}. They measured the strength of the 33.6$\, \mu$m band from model spectra for different mass-loss rates and compared those values to a selection of observed AGB stars with known mass-loss rates. The work of \citet{kemper01} showed that the selected AGB stars all have forsterite abundances below 5\%. 

We have done an expansion on the work of \citet{kemper01}. We used two spectral features of forsterite: the 11.3$\, \mu$m band as well as the 33.6$\, \mu$m band. In Fig. \ref{totalSpec} a synthetic spectrum of an AGB star is shown in which these two features are indicated. The 33.6 $\mu$m feature is not a robust, model independent indicator of the forsterite abundance because it strongly depends on the contribution of cold, optically thin dust at large distance from the star. Its strength is affected by the details of the mass loss history which is not well constrained by the dust spectral energy distribution.

On the other hand, we find that the 11.3$\, \mu$m band is a good indicator of the forsterite abundance for AGB stars with their 9.7$\, \mu$m band of amorphous silicate in absorption. The 11.3$\, \mu$m band is a feature that shows itself as a shoulder on top of the 9.7$\, \mu$m band of amorphous silicate. For mass-loss rates higher than $\sim$\nobreak$2\times10^{-5}$ M$_{\odot}$ yr$^{-1}$ this leads to very high optical depths in the region of the feature, making it much less sensitive to the details of the mass loss history. In fact, it is sensitive to the most recent mass loss of the star. The strength of the feature does not change much as a function of mass loss, above the threshold value of about $2\times10^{-5}$ M$_{\odot}$ yr$^{-1}$ (Fig. \ref{res}). This makes the 11.3$\, \mu$m band ideal as a forsterite abundance indicator. So, for high mass loss rates we can directly compare the observed strength of the forsterite absorption at 11.3 $\mu$m to model calculations without a precise measurement of the mass loss rate, or the mass loss history. We have applied this method to six OH/IR stars observed with ISO-SWS. By measuring the strength of the 11.3$\, \mu$m bands of these observations and placing them in Fig. \ref{res}, forsterite abundances are obtained.

\section{Conclusions}
We show that applying the 33.6$\, \mu$m band is practically impossible without knowledge of the parameters that influence the emission of the cold dust in the outskirts of the dust shell of the AGB star. Our analysis shows that the 11.3$\, \mu$m forsterite band is a robust indicator for the forsterite abundance of the current mass-loss period for AGB stars with an optically thick dust shell. We used the 11.3$\, \mu$m band as an indicator of the forsterite abundance for six AGB stars. This showed that these objects can have low (below 2\%) but also very high forsterite abundances (12\% and possibly higher), meaning that AGB stars can have a forsterite abundance higher than 5\% by mass, which was the maximum value found by \citet{kemper01}. It also means that the discrepancy between the amount of crystalline material in the ISM and that in AGB stars could even be higher.
If the found forsterite abundances for the six sources in our study are representative for all (high mass loss) AGB stars, the forsterite abundance in AGB stars alone would be higher than the crystalline abundance (upper limit of 2.2\%) found by \citet{kemperErr05} for the ISM. To find out if the abundances found for these stars are typical for AGB stars, more spectra have to be analyzed. Determining the amount of forsterite that is produced by high mass-loss rate AGB stars will help to understand the discrepancy between the produced abundance in AGB stars and the abundance found in the ISM.

\acknowledgements B.L. de Vries acknowledges support from the Fund for Scientific Research of Flanders (FWO) under grant number 6.0470.07.

\bibliography{devriest}

\end{document}